\documentstyle[graphicx,referee]{mn}

\newif\ifAMStwofonts

\ifoldfss
  \ifCUPmtlplainloaded \else
    \NewTextAlphabet{textbfit} {cmbxti10} {}
    \NewTextAlphabet{textbfss} {cmssbx10} {}
    \NewMathAlphabet{mathbfit} {cmbxti10} {} 
    \NewMathAlphabet{mathbfss} {cmssbx10} {} 
  \fi
  \ifAMStwofonts
    \ifCUPmtlplainloaded \else
      \NewSymbolFont{upmath} {eurm10}
      \NewSymbolFont{AMSa} {msam10}
      \NewMathSymbol{\upi}     {0}{upmath}{19}
      \NewMathSymbol{\umu}     {0}{upmath}{16}
      \NewMathSymbol{\upartial}{0}{upmath}{40}
      \NewMathSymbol{\leqslant}{3}{AMSa}{36}
      \NewMathSymbol{\geqslant}{3}{AMSa}{3E}

      \let\leq=\leqslant 
       \let\ge=\geqslant
    \fi
  \fi
\fi 

\ifnfssone
  \newmathalphabet{\mathit}
  \addtoversion{normal}{\mathit}{cmr}{m}{it}
  \addtoversion{bold}{\mathit}{cmr}{bx}{it}
  \newmathalphabet{\mathbfit} 
  \addtoversion{normal}{\mathbfit}{cmr}{bx}{it}
  \addtoversion{bold}{\mathbfit}{cmr}{bx}{it}
  \newmathalphabet{\mathbfss} 
  \addtoversion{normal}{\mathbfss}{cmss}{bx}{n}
  \addtoversion{bold}{\mathbfss}{cmss}{bx}{n}
  \ifAMStwofonts
    \ifCUPmtlplainloaded \else
      \UseAMStwoboldmath
      \makeatletter
      \new@mathgroup\upmath@group
      \define@mathgroup\mv@normal\upmath@group{eur}{m}{n}
      \define@mathgroup\mv@bold\upmath@group{eur}{b}{n}
      \edef\UPM{\hexnumber\upmath@group}
      \new@mathgroup\amsa@group
      \define@mathgroup\mv@normal\amsa@group{msa}{m}{n}
      \define@mathgroup\mv@bold\amsa@group{msa}{m}{n}
      \edef\AMSa{\hexnumber\amsa@group}
      \makeatother
      \mathchardef\upi="0\UPM19
      \mathchardef\umu="0\UPM16
      \mathchardef\upartial="0\UPM40
      \mathchardef\leqslant="3\AMSa36
      \mathchardef\geqslant="3\AMSa3E

      \let\leq=\leqslant 
       \let\ge=\geqslant
    \fi
  \fi
\fi 

\ifnfsstwo
  \DeclareMathAlphabet{\mathbfit}{OT1}{cmr}{bx}{it}
  \SetMathAlphabet\mathbfit{bold}{OT1}{cmr}{bx}{it}
  \DeclareMathAlphabet{\mathbfss}{OT1}{cmss}{bx}{n}
  \SetMathAlphabet\mathbfss{bold}{OT1}{cmss}{bx}{n}
  \ifAMStwofonts
    \ifCUPmtlplainloaded \else
      \DeclareSymbolFont{UPM}{U}{eur}{m}{n}
      \SetSymbolFont{UPM}{bold}{U}{eur}{b}{n}
      \DeclareSymbolFont{AMSa}{U}{msa}{m}{n}
      \DeclareMathSymbol{\upi}{0}{UPM}{"19}
      \DeclareMathSymbol{\umu}{0}{UPM}{"16}
      \DeclareMathSymbol{\upartial}{0}{UPM}{"40}
      \DeclareMathSymbol{\leqslant}{3}{AMSa}{"36}
      \DeclareMathSymbol{\geqslant}{3}{AMSa}{"3E}

      \let\leq=\leqslant 
       \let\ge=\geqslant
    \fi
  \fi
\fi 

\ifCUPmtlplainloaded \else
  \ifAMStwofonts \else 
    \def\upi{\pi}
    \def\umu{\mu}
    \def\upartial{\partial}
  \fi
\fi

\title{The non-Gaussian tail of cosmic-shear statistics}
\author[Guido Kruse \& Peter Schneider]
       {Guido Kruse \& Peter Schneider\\
        Max-Planck-Institut f\"ur Astrophysik, Postfach 1523, D-85740,
Garching, Germany}
\date{Accepted 1988 December 15.
      Received 1988 December 14;
      in original form 1988 October 11}

\pagerange{\pageref{firstpage}--\pageref{lastpage}}
\pubyear{1998}

\begin{document}

\maketitle

\label{firstpage}

\begin{abstract}
Due to gravitational instability, an initially Gaussian density field
develops non-Gaussian features as the Universe evolves. The most
prominent non-Gaussian features are massive haloes, visible as
clusters of galaxies. The distortion of high-redshift galaxy images
due to the tidal gravitational field of the large-scale matter
distribution, called cosmic shear, can be used to investigate the
statistical properties of the LSS. In particular, non-Gaussian
properties of the LSS will lead to a non-Gaussian distribution of
cosmic-shear statistics. The aperture mass ($M_{\rm ap}$) statistics, recently
introduced as a measure for cosmic shear, is particularly well suited
for measuring these non-Gaussian properties. In this paper we
calculate the highly non-Gaussian tail of the aperture mass
probability distribution, assuming Press-Schechter theory for the halo
abundance and the `universal' density profile of haloes as obtained
from numerical simulations. We find that for values of $M_{\rm ap}$
much larger than its dispersion, this probability distribution is
closely approximated by an exponential, rather than a Gaussian. We
determine the amplitude and shape of this exponential for various
cosmological models and aperture sizes, and show that wide-field
imaging surveys can be used to distinguish between some of the
currently most popular cosmogonies. Our study here is complementary to
earlier cosmic-shear investigations which focussed more on two-point
statistical properties. 
\end{abstract}

\begin{keywords}
galaxies: clusters: general - cosmology: theory - dark matter -
gravitational lensing
\end{keywords}

\section{Introduction}

The deflection of light due to the gravitational field of matter
inhomogeneities is observable through the distortion of images of
background galaxies. In the case of weak gravitational fields, i.e.,
in the absence of strong gravitational lensing effects like giant
arcs, the images of a population of background sources with known
intrinsic ellipticity distribution can be used to statistically
investigate weak gravitational lensing effects by measuring a net
ellipticity.

Cosmic shear -- the line-of-sight integrated tidal gravitational field
-- reflects the statistical properties of the density fluctuation
field (Gunn
1967, Blandford \& Jaroszynski 1981).  A quantitative description of
this connection is given by the two-point correlation function of
galaxy-image ellipticities, or by the rms-ellipticity within a
(circular) aperture, which was investigated by several authors using
the linear and nonlinear power spectrum of density fluctuations in
different cosmologies (see the recent review by Mellier 1998, and
references therein).

Schneider et al. (1998; hereafter SvWJK) investigated cosmic shear
using the $M_{\rm ap}$-statistics which is a spatially filtered
version of the projected density field. They computed the dispersion
of $M_{\rm ap}$ for the linear and nonlinear power spectrum of density
fluctuations in different cosmologies for a broad range of filter
scales and showed that this quantity is a sensitive and `local'
measure of the power spectrum; in fact, as shown by Bartelmann \&
Schneider (1999), the mean-squared value of $M_{\rm ap}$ on an angular scale
$\theta$ is a very good approximation to the power spectrum $P_\kappa$
of the projected density field at a wavenumber $s\approx 4.25/\theta$.
SvWJK calculated the skewness of $M_{\rm ap}$ in the frame of
quasi-linear theory of structure growth and found that it is a
sensitive indicator for the density parameter $\Omega_0$, independent
of the normalization of the power spectrum (see also Bernardeau, van
Waerbeke \& Mellier 1997; van Waerbeke, Bernardeau \& Mellier 1999).
The skewness measures the
non-Gaussianity of the projected density field and indicates,
according to its sign, an asymmetric positive or negative tail of the
probability distribution function (PDF). The skewness of $M_{\rm ap}$,
which reflects the skewness of the three-dismensional density
fluctuations, was found to be positive, indicating an extended tail of
the PDF towards high positive values of $M_{\rm ap}$. This behaviour is
expected: the initial density contrast $\delta$, which is assumed to
be a Gaussian random field, becomes non-Gaussian during its evolution
through gravitational collapse.  That means the PDF of $\delta$
attains a cut-off near $\delta=-1$ and a broad positive tail, according
to the occurrence of underdense and overdense regions. Since $M_{\rm
ap}$ is a linear function of the projected density field, the PDF of
$M_{\rm ap}$ is closely related to that of $\delta$.  Therefore the
PDF of $M_{\rm ap}$ reflects the nonlinear evolution of $\delta$.

The projected density field $\kappa$ in general is defined as a
projection of the three-dimensional density contrast $\delta$, weighted
by a redshift-dependent factor which accounts for the lensing geometry and
the redshift distribution of the sources. The highest peaks of $\kappa$
are expected to arise from physical objects
with high three-dimensional density, i.e., collapsed haloes. In Kruse
\& Schneider (1999; hereafter KS99) we have calculated the number
density of haloes which yield a value of $M_{\rm ap}$ larger than a certain
threshold. We found that for all cosmologies considered, the number
density of haloes above a threshold corresponding to a signal-to-noise
of 5 exceeds ten per square degree,
for currently feasible deep optical imaging, 
so that a wide-field deep optical survey should detect
hundreds of such peaks, which can then be used to define a
mass-selected sample of dark matter haloes (Schneider 1996).

In this paper we continue the investigation of statistical
properties of the aperture mass by computing its cumulative probability
distribution function (CPDF) for large positive values of $M_{\rm
ap}$, assuming that all of these are caused by collapsed structures. 
Describing the density profile of the haloes by the universal mass profile
found by Navarro, Frenk \& White (1996, 1997; hereafter NFW), we can
assign an angular cross-section to each halo. If this cross-section is
integrated over the abundance of haloes as obtained from
Press-Schechter theory, one can determine the probability to measure a
value of $M_{\rm ap}$ larger than some threshold, i.e., the CPDF for
the tail of $M_{\rm ap}$. Similar to the number density calculated in KS99,
the amplitude and shape of this tail reflects the abundance of 
dark-matter haloes, which can be used as a powerful cosmological probe.
 
The rest of this paper is organized as follows: in Sect.\ 2 we briefly
review the concept of the aperture mass, as applied to the NFW-
profile. The cross-sectional area as a function of $M_{\rm ap}$, and
the CPDF, are derived in Sect.\ 3, and results are given in Sect.\
5. The number density of peaks in the two-dimensional distribution of
$M_{\rm ap}$ will be strongly affected, though in a controllable way,
by noise, mainly coming from the intrinsic ellipticity distribution of
galaxies. We consider an alternative observable for the abundance of
peaks in $M_{\rm ap}$ in Sect.\ 4. Finally, we summarize and discuss
our results in Sect.\ 6.

\def\d{{\mathrm d}}

\section{Formalism}

Following Schneider (1996), we define
the spatially filtered mass inside a circular aperture of angular
radius $\theta$ around the point 
$\mbox{\boldmath$\zeta$}$,
\begin{equation}
M_{\rm ap} (\mbox{\boldmath$\zeta$}):=\int {\rm d}^2 \vartheta 
\ \kappa(\mbox{\boldmath$\vartheta$})
\ U(\vert \mbox{\boldmath$\vartheta-\zeta$} \vert),
\label{mapt}
\end{equation}
where
the continuous weight function $U(\vartheta)$ vanishes for
$\vartheta>\theta$.
If $U(\vartheta)$ is a compensated filter function,
\begin{equation}
\int_0^{\theta} {\rm d} \vartheta \ \vartheta \ U(\vartheta)=0,
\end{equation} 
one can express $M_{\rm ap}$ in terms of the tangential shear 
$\gamma_{\rm t}(\mbox{\boldmath$\xi$}; \mbox{\boldmath$\zeta$})$ 
at position 
$\mbox{\boldmath$\xi + \zeta$}$
relative to the point $\mbox{\boldmath$\zeta$}$
\begin{equation}
M_{\rm ap} (\mbox{\boldmath$\zeta$})=\int {\rm d}^{2} \xi \ 
\gamma_{\rm t} (\mbox{\boldmath$\xi$}; \mbox{\boldmath$ \zeta$}) 
\ Q(\vert \mbox{\boldmath$\xi$} \vert),
\label{mapshear}
\end{equation}
where
\begin{equation}
\gamma_{\rm t}(\mbox{\boldmath$\xi$}; \mbox{\boldmath$\zeta$}) 
= -{\rm Re}(\gamma(\mbox{\boldmath$\xi+\zeta$}) e^{-2i \phi})
\label{tanshear}
\end{equation}
is the tangential component of the shear at relative position 
$\mbox{\boldmath$\xi$}=(\xi \ {\rm cos} \ \phi, 
\xi \ {\rm sin} \ \phi)$.
The function $Q$ is related to $U$ by
\begin{equation}
Q(\vartheta) = \frac{2}{\vartheta^2} \ \int_0^{\vartheta}
{\rm d} \vartheta^{\prime} \ \vartheta^{\prime} \ U(\vartheta^
{\prime})
\ - U(\vartheta) .
\end{equation}
We use a filter function from the family given in SvWJK,
specifically we choose that with $l=1$. 
Then writing $U(\vartheta)=u(\vartheta/
\theta)/\theta^2$, and $Q(\vartheta)=
q(\vartheta/\theta)/\theta^2$, 
\begin{equation}	
u(x)=\frac{9}{\pi} (1-x^2) \left( \frac{1}{3}-x^2 \right), 
\end{equation}
and
\begin{equation}	
q(x)=\frac{6}{\pi} x^2(1-x^2),
\end{equation}
with $u(x)=0$ and $q(x)=0$ for $x>1$.
We will describe the mass density of dark matter haloes with the
universal density profile introduced by NFW,
\begin{equation}
\rho (r)= \frac{3 H_0^2}{8 \pi G} \ (1+z)^3 \ \frac{\Omega_{\rm d}}{\Omega
(z)} \ \frac{\delta_c}{r/r_{\rm s} (1+r/r_{\rm s})^2},  
\end{equation}
with
\begin{equation}
\Omega(z)=\frac{\Omega_{\rm d}}{a+\Omega_{\rm d} (1-a)+\Omega_{\rm v} 
(a^3-a)}, \ a=\frac{1}{1+z}.  
\end{equation}
$\Omega_{\rm d}$ and $\Omega_{\rm v}$ denote the present-day density
parameters in dust and in vacuum energy, respectively.
Haloes identified at redshift $z$ with mass $M$ are described by the
characteristic density $\delta_{\rm c}$ and the scaling radius 
$r_{\rm s}=r_{200}/c$ where $c$ 
is the concentration parameter (which is a function of
$\delta_{\rm c}$) and $r_{200}$ is the virial radius, defined such
that a sphere with radius $r_{200}$ has a mean interior density 
of $200 \ \rho_{\rm crit}$ and contains the halo mass $M_{200}$.
We compute the parameters which specify
the density profile according to the description in NFW using the fitting
formulae given there. 
The surface mass density of the NFW-profile is given by 
(see Bartelmann 1996)	
\begin{equation}
\Sigma (\vartheta)=  \frac{6 H_0^2}{8 \pi G} \ (1+z)^3 \ 
\frac{\Omega_{\rm d}}{\Omega
(z)} \ r_{\rm s} \ \delta_{\rm c} \
f \left( \frac{\vartheta}{\theta_{\rm s}} \right),  
\end{equation}
with
\begin{eqnarray}
f(x)&=&\frac{1}{x^2-1}  \nonumber \\
&\times \ \biggl\{&^{\displaystyle{
1-\frac{2}{\sqrt{1-x^2}} \ \mbox{arctanh} \ 
\sqrt{\frac{1-x}{1+x}}, \ \mbox{for} \ x < 1}}
_{\displaystyle{1-\frac{2}{\sqrt{x^2-1}} \ \mbox{arctan} \ 
\sqrt{\frac{x-1}{1+x}}, \ \mbox{for} \ x > 1}},
\end{eqnarray}
and $\theta_{\rm s}= r_{\rm s}/D_{\rm d}$.
$D_{\rm d}$ is the angular diameter distance to the lens.
Introducing the critical surface mass density
\begin{equation}
\Sigma_{\rm cr}= \frac{c^2}{4 \pi G} \ \frac{D_{\rm s}}
{D_{\rm d} D_{\rm ds}}, 
\label{geom}
\end{equation}
with $D_{\rm s}$ and $D_{\rm ds}$ being the angular
diameter distances
to the source and that from the lens to the source,
we can define the dimensionless surface density (convergence)
which is a function of source redshift
\begin{equation}
\kappa (\vartheta,z_{\rm d},z_{\rm s}) = \frac{\Sigma (\vartheta)}
{\Sigma_{\rm cr}} =
\kappa_0 \ f \left( \frac{\vartheta}{\theta_{\rm s}} \right),
\end{equation}
with
\begin{equation}
\kappa_0=3 \ (1+z)^3 \ \frac{\Omega_{\rm d}}{\Omega(z)}\ r_{\rm s} \
\frac{H_0^2}{c^2} \ \delta_c \ \frac{D_{\rm d} D_{\rm ds}}{D_{\rm s}}. 
\end{equation}
We assume a normalized source redshift distribution of the form
\begin{equation}	
p_z (z)= \frac{\beta}{z_0^3 \ \Gamma \left(\frac{3}{\beta} \right)}
\ z^2 \ \mbox{exp}(-[z/z_{0}]^{\beta}), 
\label{sources}
\end{equation}	
(see Brainerd et al. 1996).
The mean redshift of this distribution is proportional to 
$z_{0}$ and depends on the parameter $\beta$ which describes
how quickly the distribution falls off towards higher redshifts.
We will use the values $\beta=1.5$ and $z_0=1$. For these values
the mean redshift $\langle z \rangle$ is given by
$\langle z \rangle =1.505 \ z_{0}$.
With the distribution (\ref{sources}) we define 
a source distance-averaged surface density 
\begin{equation}	
\kappa (\vartheta,z_{\rm d}) = \int {\rm d}z_{\rm s} \ p_z 
(z_{\rm s}) \ \kappa (\vartheta,z_{\rm d},z_{\rm s})
= \bar \kappa_0 \ f \left( \frac{\vartheta}{\theta_{\rm s}} \right), 
\label{ka}
\end{equation}
with $\bar \kappa_0 = \int {\rm d} z_{\rm s} \ p_z (z_{\rm s}) \
\kappa_0$. 
Inserting (\ref{ka}) in (\ref{mapt}) we get, after introducing polar
coordinates for $\mbox{\boldmath$\vartheta$}$ and setting
$\mbox{\boldmath$\zeta$}=(\zeta,0)$, which is the radial
distance of the aperture centre from the halo center,
\begin{eqnarray}
&&M_{\rm ap} (\zeta)= \int_{{\rm max}(0,\zeta-\theta)}
^{\zeta+\theta}
{\rm d} \vartheta \ \vartheta \ \kappa(\vartheta) \ \nonumber \\
&& \times \int_{-\phi_{\rm m}}^{\phi_{\rm m}} {\rm d} \phi \ 
 U\left(\sqrt{\vartheta^2+\zeta^2-2 \vartheta \zeta {\rm cos}
(2 \phi)}\right),
\label{map}
\end{eqnarray}
where $\phi_{\rm m}= {\rm min} \left( \pi, {\rm arccos}
\left(\frac{\vartheta^2+\zeta^2-\theta^2}{2 \vartheta \zeta}\right)
\right)$. For the filter function chosen above , the
$\phi$-integration can be performed analytically. 

\begin{figure}
\center{\includegraphics[
        width=\columnwidth,
       draft=false,
        ]{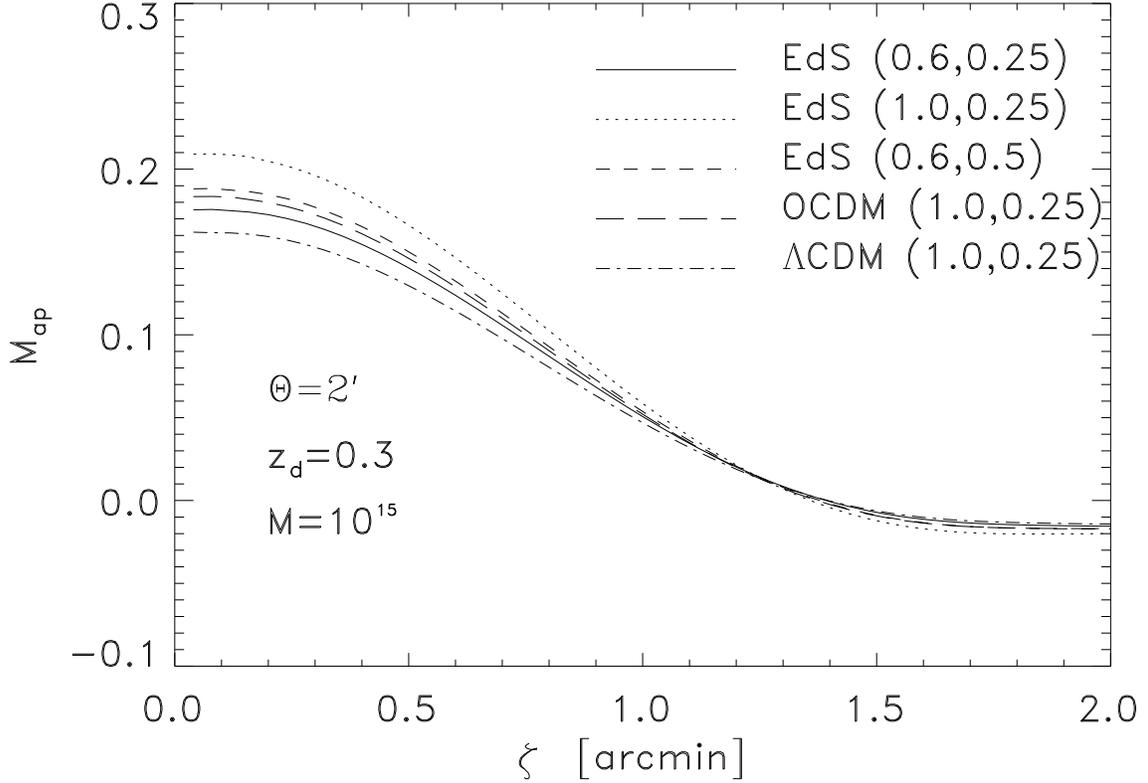}}
\caption{The aperture mass, as defined in (\ref{map}), as a function
of the radial distance $\zeta$ computed for
five different cosmologies as indicated by the line types. The numbers
in parentheses are the normalization $\sigma_8$ annd the shape
parameter $\Gamma$. The halo mass is $10^{15} M_{\odot}/h$ and its 
redshift $z_{\rm d}=0.3$. The filter radius is $\theta=2$ arcmin.
\label{map_zeta}}
\end{figure}

In Figure \ref{map_zeta} we plot the aperture mass of a halo with
mass $M=10^{15} M_{\odot}/h$ at redshift $z_{\rm d}=0.3$ as a function
of the radial distance $\zeta$, using a filter
scale of $\theta=2'$, for five cosmological models. 
If we fix the parameters $(M,z_{\rm d},\theta)$, the aperture mass
is a monotonically decreasing function of $\zeta$
between $\zeta=0$ and the first root of $M_{\rm ap}$.

\section{The tail of $M_{\rm ap}$-statistics}

In this section we calculate the CPDF $P(>M_{\rm ap},\theta)$, i.e.,
the probability to find an aperture mass larger than $M_{\rm
ap}$ using a filter with radius $\theta$. We concentrate on values 
of $M_{\rm ap}$ which are much larger than the rms value of $M_{\rm
ap}$, i.e., we consider only the far tail of the probability
distribution. We may assume that such high values of $M_{\rm ap}$
are caused exclusively by collapsed haloes. Thus, from the properties
of such haloes, together with their abundance, we can then determine
the CPDF.

Assuming a halo characterized by its mass $M$ and redshift $z_{\rm
d}$, we can invert the function (\ref{map}) for a given value of
$M_{\rm ap}$ and a fixed filter radius $\theta$ (see
Fig. \ref{map_zeta}).  As a result we get the separation
$\zeta=\zeta(M_{\rm ap}, \theta, M, z_{\rm d})$. Owing to the
monotonic behaviour of $M_{\rm ap}$ between $\zeta=0$ and its first
root, separations smaller than the one obtained by 
inversion correspond to aperture masses larger than the threshold
$M_{\rm ap}$.  Therefore, $\zeta$ defines for each halo an angular cross
section
\begin{equation}
\sigma(M_{\rm ap},\theta,M,z_{\rm d}) := \pi \ \zeta^2 (M_{\rm
ap},\theta,M,z_{\rm d}),
\label{cro}
\end{equation}
which represents the halo target area for detecting a weak lensing
signal with an aperture mass larger than $M_{\rm ap}$. The
cross section (\ref{cro}) is non-zero if the aperture mass measured in
the halo centre is larger than the threshold $M_{\rm ap}$ (see
Fig. \ref{map_zeta}).

The CPDF is now obtained by summing up the cross sections of all
haloes within a unit solid angle; this yields
\begin{eqnarray}
&&P(>M_{\rm ap},\theta) = \frac{c}{H_0} \ \int \d z_{\rm d} \ \frac{
(1+z_{\rm d})^2}{E(z_{\rm d})} \ D_{\rm d}^2 (z_{\rm d}) \ \nonumber \\
&&\times \int {\rm d} M \ N_{\rm halo}(M,z_{\rm d}) \ 
\sigma(M_{\rm ap}, \theta,M,z_{\rm d}),
\label{prob2}
\end{eqnarray}	
with
\begin{equation}	
E(z_{\rm d})= \sqrt{\Omega_{\rm d} (1+z_{\rm d})^3+ (1-\Omega_{\rm d}
-\Omega_{\rm v})(1+z_{\rm d})^2+\Omega_{\rm v}}.
\end{equation}
$N_{\rm halo}(M,z_{\rm d}) \ {\rm d} M$ is the comoving number density
of haloes with mass within ${\rm d} M$ about $M$ at redshift $z_{\rm d}$.
We assume that $N_{\rm halo}(M,z_{\rm d})$ can be obtained from Press
and Schechter (1974) theory (see, e.g. Lacey $\&$ Cole 1993,
1994). The $M$-integral in (\ref{prob2}) extends from a lower
threshold $M_{\rm t}$ to 
infinity, where $\sigma$ vanishes for $M \leq M_{\rm t}$ (see KS99).

\section{Number density of haloes, selected by $M_{\rm ap}$ and size}

\begin{figure}
\center{\includegraphics[
        width=\columnwidth,
       draft=false,
        ]{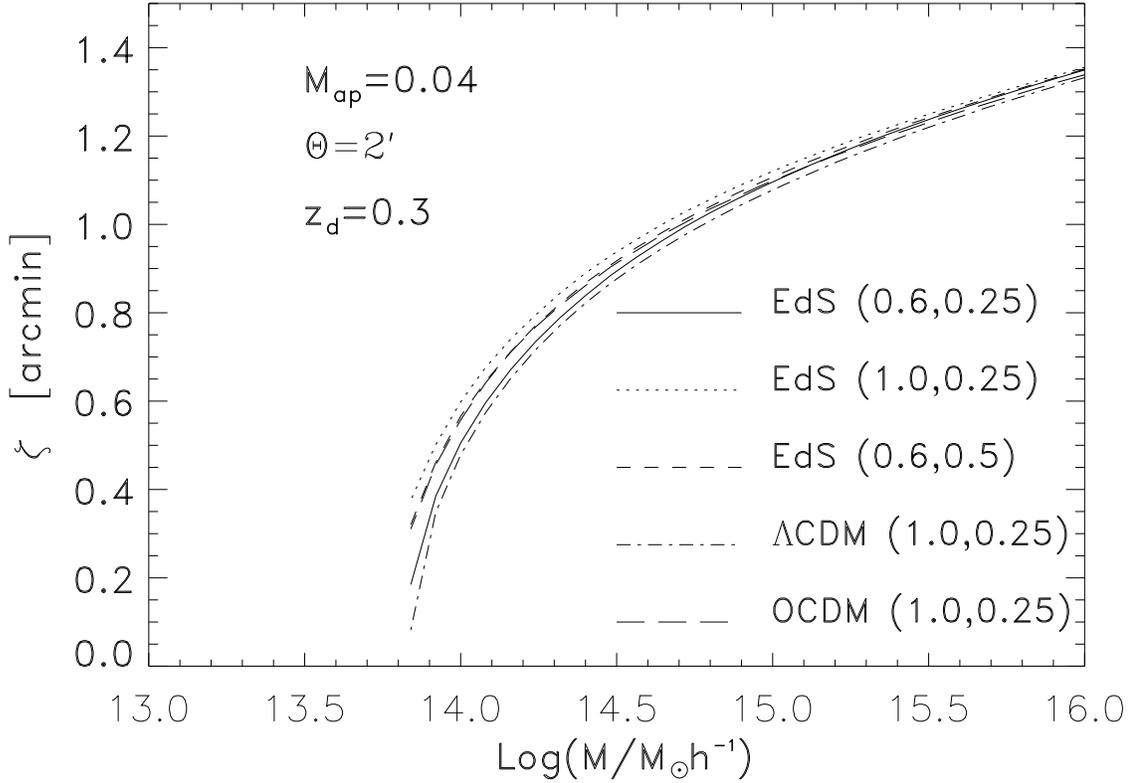}}
\caption{The cross section radius obtained from inverting (\ref{map})
as a function of the lens mass for the same cosmological models
as in Figure \ref{map_zeta}. 
(\ref{map}) is inverted for the aperture mass $M_{\rm ap}=0.04$,
a filter radius $\theta=2$ arcmin and 
a halo redshift $z_{\rm d}=0.3$.
The cross section
of haloes characterized by the fixed parameters written in the
panel is zero for halo masses smaller than $M_{\rm t} 
\sim 10^{13.8} M_{\odot}/h$.
\label{zeta_m}}
\end{figure}

In KS99, we have calculated the number density of haloes 
$N(>M_{\rm ap},\theta)$ with an aperture mass $> M_{\rm ap}$, using
the same physical model as described above. Considering $M_{\rm ap}
(\theta)$ as a two-dimensional map, $N(>M_{\rm ap},\theta)$ yields, at
fixed filter scale $\theta$, the number density of peaks in this map
with amplitude $> M_{\rm ap}$. Such peaks may be generated by noise
where the major contribution comes from the intrinsic ellipticity
distribution of galaxies. Intuitively, one might expect that high
peaks are less effected by noise than smaller ones; this motivates us
to consider the number density of haloes with aperture mass $> M_{\rm
ap}$ and cross-sectional area $>\sigma=\pi \zeta_{\rm t}^2$,
which is again obtained from summing over all haloes per unit
solid angle,
\begin{eqnarray}
&&N(> M_{\rm ap}, > \zeta_{\rm t}, \theta)= 
\frac{c}{H_0} \ \int \d z_{\rm d} \
\frac{(1+z_{\rm d})^2}{E(z_{\rm d})} \ D_{\rm d}^2 (z_{\rm d}) \ 
\nonumber\\
&&\times \int \d M \ N_{\rm halo}(M,z_{\rm d}) \ 
{\rm H}[\zeta(M_{\rm ap},\theta,M,z_{\rm d})-\zeta_{\rm t}],
\label{numbercro}
\end{eqnarray}
where ${\rm H}$ is the Heaviside step function. The integrand is
non-zero only for $M>M_{\rm t}$, where $M_{\rm t}=M_{\rm t}
(\zeta_{\rm t}, z_{\rm d} ,M_{\rm ap},\theta)$ is the mass obtained by
inversion of the function shown in Fig. \ref{zeta_m}.  For $\zeta_{\rm
t}=0$, $N(> M_{\rm ap}, > \zeta_{\rm t}, \theta)= N(> M_{\rm
ap},\theta)$.  The value of $\zeta_{\rm t}$ for which $N(> M_{\rm ap},
> \zeta_{\rm t}, \theta)$ decreases to about $1/2$ of $N(> M_{\rm
ap},\theta)$ yields the characteristic size of peaks corresponding to
a given $M_{\rm ap}$; this value is expected to be $< \theta$. E.g.,
for the EdS(0.6,0.25) cosmology we obtain $N(> 0.04, > 0.45', 2')=
4.6$ which can be compared to $N(> 0.04, 2')= 9.4$ computed in
KS99. For this example the characteristic size of a peak is roughly
$1/4$ of the filter radius.

\section{results}

In this section we consider $P(>M_{\rm ap},\theta)$ for different
cosmological models in the regime where the PDF of $M_{\rm ap}$
describes the non-linear evolution of the density field.
Furthermore, we use the
observable (\ref{numbercro}) to get additional constraints for
the cosmological parameters. We perform our calculations for
the same five cosmological models as shown in Figure \ref{map_zeta}.  
For three of them, the power spectrum is approximately cluster
normalized, which corresponds to $\sigma_8 \approx 0.6$ for an
Einstein-de Sitter universe (EdS, $\Omega_{\rm d} = 1$, 
$\Omega_{\rm v} = 0$) and $\sigma_8 = 1$ for both an open universe
(OCDM, $\Omega_{\rm d} = 0.3$, $\Omega_{\rm v} = 0$) and a spatially
flat universe with cosmological constant ($\Lambda$CDM, $\Omega_{\rm d}
= 0.3$, $\Omega_{\rm v} = 0.7$). For these models we use 
the shape parameter $\Gamma = 0.25$ which
yields the best fit to the observed two-point correlation function
of galaxies (Efstathiou 1996). The remaining two 
EdS models
have higher normalization ($\sigma_8 = 1$, approximately corresponding
to COBE normalization) or a different shape parameter ($\Gamma =
0.5$).

\begin{figure*}
\center{\includegraphics[
       draft=false,
        ]{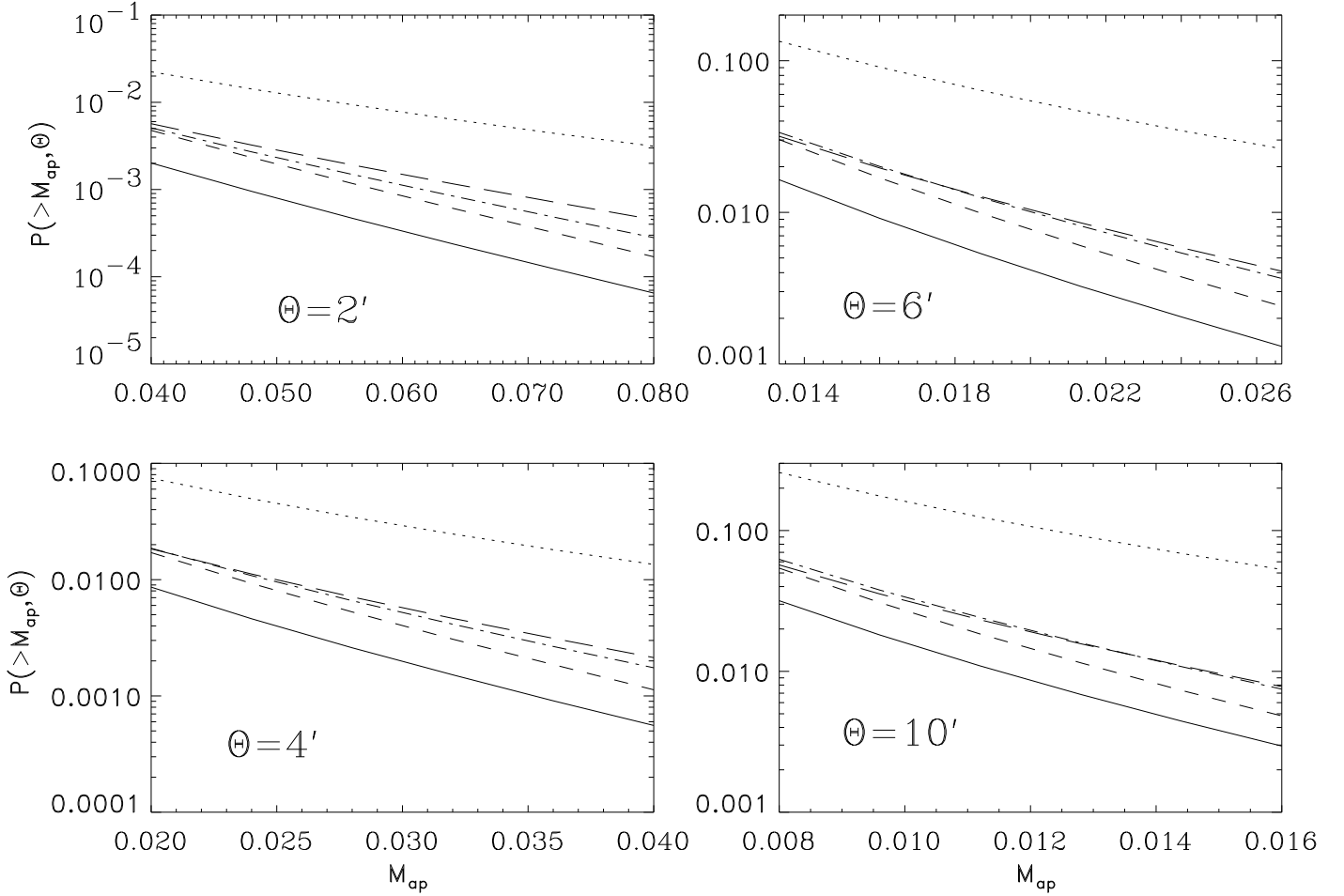}}
\caption{The probability as defined in (\ref{prob2}) for the filter
radii $\theta=2,4,6,10$ arcmin. In each panel we plot the same
cosmological models as indicated in Figs. 1 and 2. The aperture mass
range is defined by $[M_0,2 M_0]$, where $M_0=5 \cdot \sigma_{\rm c}
(\theta)$  [see (\ref{sn}) and (\ref{var})].
\label{multi_prob}}
\end{figure*}

In the absence of lensing, the expectation value of $M_{\rm ap}$
vanishes, and its dispersion depends on the intrinsic ellipticity
distibution, as calculated in Schneider (1996),
\begin{equation}
\sigma_{\rm c} (\theta) = 
0.016 \left( \frac{n}{
30 \ {\rm arcmin}^{-2}} \right)^{-1/2}
\left( \frac{\sigma_{\epsilon}}{0.2} \right)
\left( \frac{\theta}{1'}  \right)^{-1}.
\label{var}
\end{equation}
We take $n=30$ arcmin$^{-2}$ and $\sigma_{\epsilon}=0.2$
for the number density of background galaxies and their intrinsic
ellipticity distribution, respectively, as representative values in
the following. The signal-to-noise ratio of $M_{\rm ap}$ is then
defined as
\begin{equation}
S_{\rm c} (\theta) = \frac{M_{\rm ap} (\theta)}{\sigma_{\rm c}
(\theta)}. 
\label{sn}
\end{equation}

In Fig. \ref{multi_prob} we plot the probability (\ref{prob2}) for four
different filter scales for the cosmologies introduced in the
beginning of this section, for values of $M_{\rm ap}$ between $M_0$
and $2M_0$, where the value of $M_0$ was chosen to correspond to a
signal-to-noise ratio of five, $M_0=5 \sigma_{\rm c} (\theta)$.
According to this figure the probability
$P(>M_{\rm ap},\theta)$ can be well approximated by an exponential,
\begin{equation}
P(>M_{\rm ap},\theta) = p_0 \ {\rm exp} \left[- \frac{(M_{\rm ap}-
M_0)}{c} \right],
\label{fit}
\end{equation}
where $p_0$ and $c$ are fit parameters.  From (\ref{fit}) it is clear
that $p_0$ is the probability to find a value of $M_{\rm ap}$ larger
than $M_0$. We determined $p_0$ and $c$ simply by assuming that the
logarithm of $P(>M_{\rm ap},\theta)$ in the interval $[M_0,2 M_0]$
follows a straight line.  The maximal relative deviation between the
probability (\ref{prob2}) and its approximation (\ref{fit}) in
Fig. \ref{multi_prob} is about 3 percent.

\begin{figure*}
\center{\includegraphics[
       draft=false,
        ]{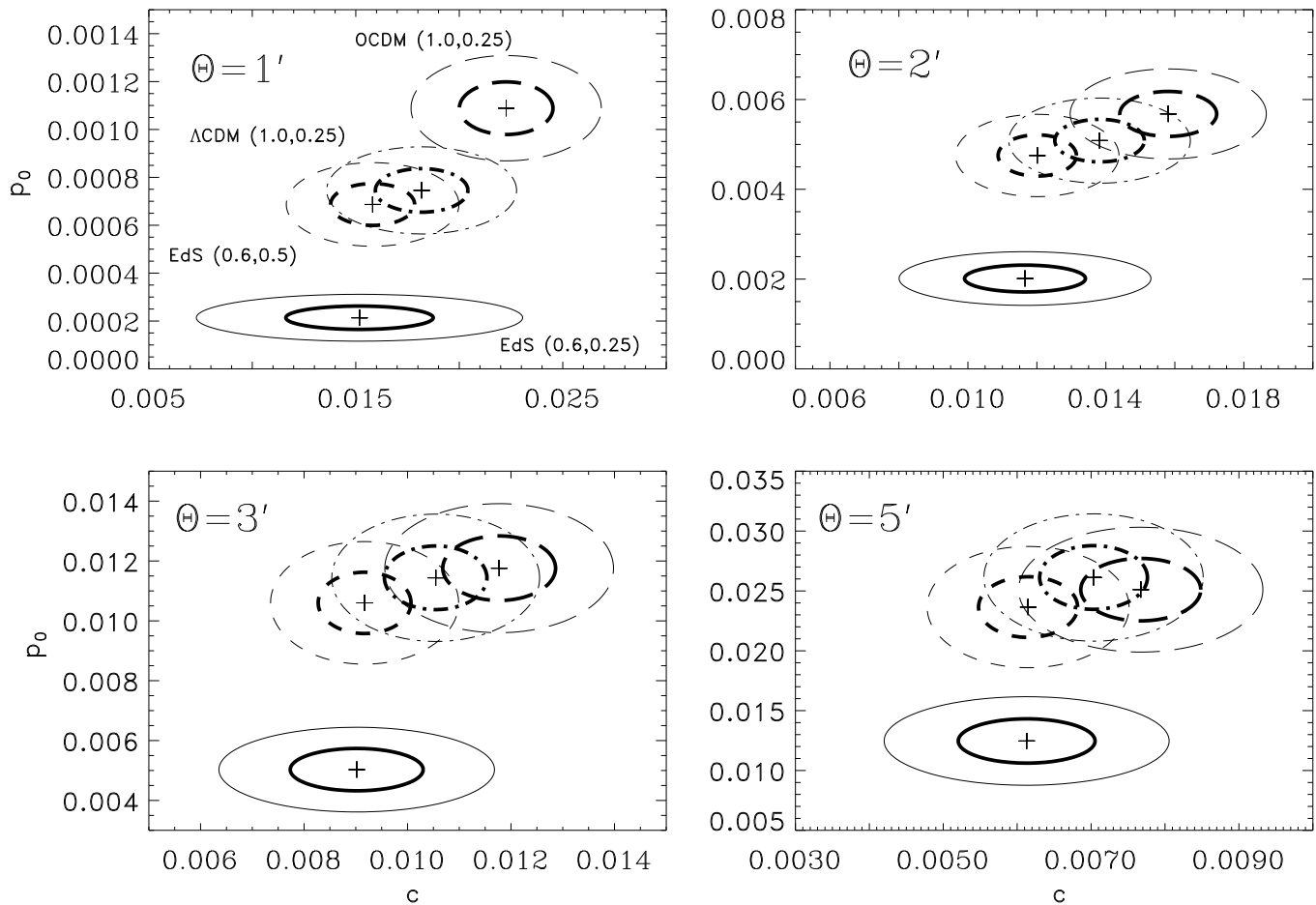}}
\caption{The fit parameters of the exponential (\ref{fit})
denoted by crosses computed as described
in the text together with their 1-$\sigma$ error ellipses which are
defined by the
dispersions (\ref{disp_c}) and (\ref{disp_p0}). For all panels the
signal-to-noise ratio threshold $M_0/\sigma_{\rm c}$
is 5. Haloes with redshift in $z_{\rm d} \in
[0,1]$ are considered. We use the source redshift distribution
(\ref{sources}) with $\beta=1.5$ and $z_0=1$.
The thin and the thick curves describe a
25 deg$^2$ and a 100 deg$^2$ survey, respectively. The cosmological
models are indicated in the upper left panel.
The different panels are obtained by varying the filter scale, 
$\theta=1,2,3,5$ arcmin. We do not plot the EdS(1.0,0.25) models because it
is well separated from the other cosmologies ($p_0$ is about a factor
of 5 larger compared to the probabilities of the remaining model).
\label{multi_2_survey}}
\end{figure*}

\begin{figure*}
\center{\includegraphics[
       draft=false,
        ]{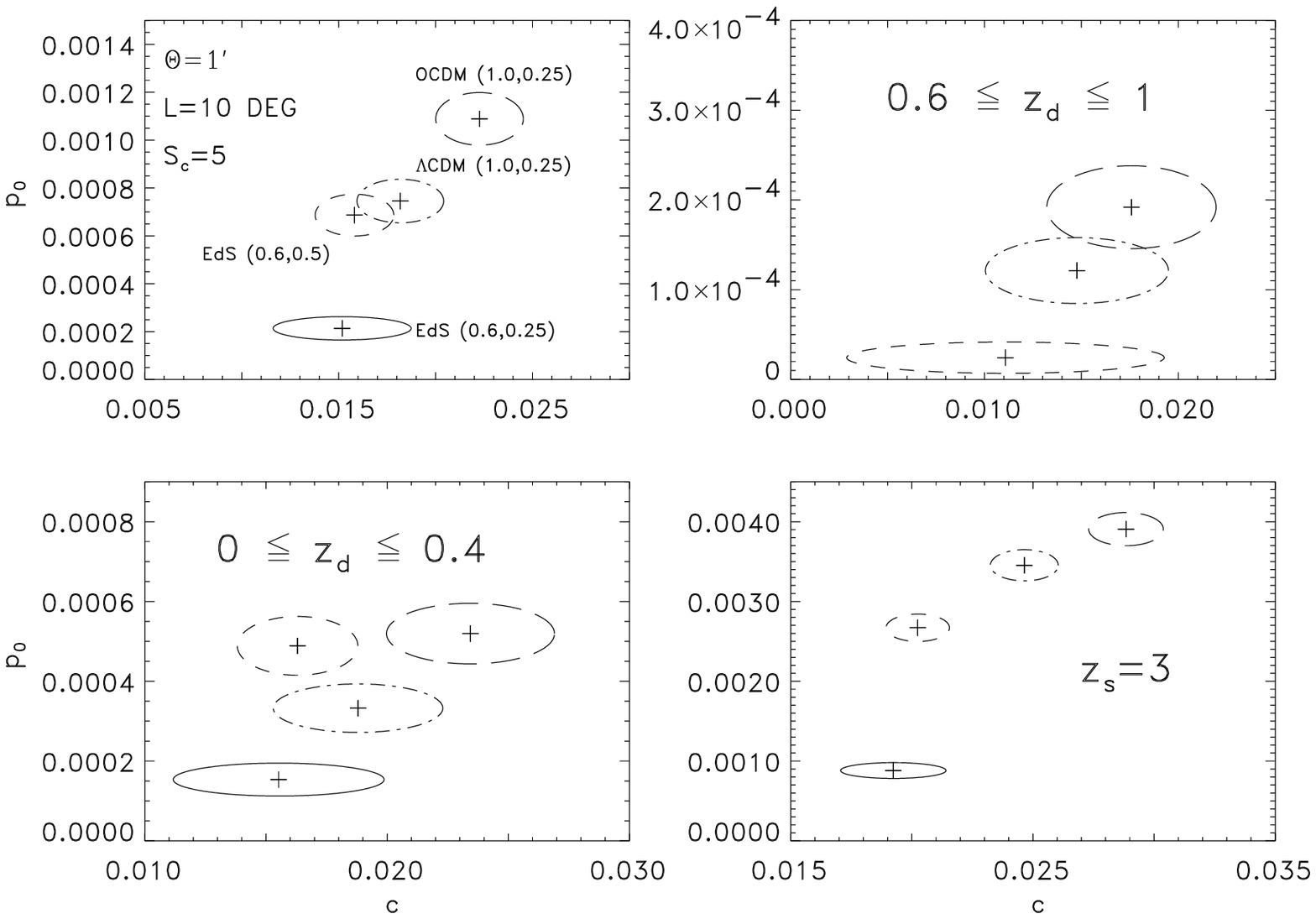}
\caption{The {top left} plot is the same as in
Fig. \ref{multi_2_survey} for $L=10$ deg. In the {top right} plot
all parameters are unchanged but only halo redshifts $z_{\rm d} \in
[0.6,1]$ are considered. 
The {bottom left} plot is the same as
the {top right} one but halo redshifts are from the interval  
$z_{\rm d} \in [0,0.4]$.
In the {bottom right} plot all sources are located 
at redshift $z_{\rm s}=3$; the other parameters are the same as in 
the {top left} plot. As in Fig. \ref{multi_2_survey} we do not
plot the high normalized EdS model which again is well separated from
the other models. In the top right panel
the value of $p_0$ for the 
EdS(0.6,0.25) model is so small that the normalization constant $f$
differs from unity appreciably; we have therefore not included this
model in the figure. The cosmological models
are indicated in the upper left panel}.
\label{multi_mix_new}}
\end{figure*}

If we compute the probability according to the approximation
(\ref{fit}), all information on cosmology is contained in the
parameters $p_0$ and $c$. If we attempt to constrain 
the cosmological parameter set, we have to compute dispersions for the
fit parameters. In the following we employ a maximum likelihood
analysis to derive confidence levels in the parameter space $\{c,p_0\}$.

In order to obtain a likelihood function we have to specify a
probability distribution for finding a particular set of values 
$\{ M_{\rm ap}^{\rm i} \}$, ${\rm i} \in
[1,N_{\rm f}]$ of $N_{\rm f}$ statistically independent aperture mass values.
As an example, we
shall assume that we have an image with side length $L$. SvWJK showed
that about $N_{\rm f}= (L/(2 \theta))^2$ statistically independent
fields can be placed on the image. This means that we consider two
fields to be statistically independent if their angular separation is
about two filter radii since then the correlation coefficient of the two
fields is below 1 percent (see SvWJK).
We suppose that this sample contains $N_>$ values of $M_{\rm ap}$
which are above the threshold $M_0$. This subsample consists of 
aperture mass values $M_{\rm ap}^{\rm j}>M_0$, ${\rm j} \in [1,N_>]$. 
Now we may ask for the
probability to find $N_> \in [1,N_{\rm f}]$ values above the
threshold if we have a sample of $N_{\rm f}$ fields. For each of the
$N_{\rm f}$ aperture mass measurements the event $M_{\rm ap} > M_0$
is supposed to occur with probability $p_0$. 
Obviously this random process can be described by a binomial
distribution,
\begin{equation}
P(N_>| N_{\rm f},p_0) = { N_{\rm f} \choose N_> } \ p_0^{N_>} \
(1-p_0)^{N_{\rm f}-N_>}.
\label{bino}
\end{equation}
Of course, we can hope to estimate the two parameters $p_0$ and
$c$ only for observations for which $N_> \gg 1$; in particular, if
$N_> =0$, the parameter $c$ will be completely undetermined. Thus,
we want to consider only realizations of (\ref{bino}) with 
$N_> \ge 1$, and therefore renormalize $P(N_>| N_{\rm
f},p_0)$,  
\begin{equation}
f \sum_{N_>=1}^{N_{\rm f}} \ P(N_>| N_{\rm f},p_0) = 1,
\end{equation}
where
\begin{equation}
f = \frac{1}{1-(1-p_0)^{N_{\rm f}}}.
\label{f}
\end{equation}
In addition, we consider the probability distribution of
$M_{\rm ap}$ itself for values above the threshold $M_0$. The
normalized PDF of $M_{\rm ap} \ge M_0$ can be 
obtained from (\ref{fit}) by
\begin{eqnarray}
\hat p(M_{\rm ap})&&=\frac{1}{p_0} \ \left|
\frac{{\rm d}}{{\rm d} M_{\rm ap}}
P(>M_{\rm ap},\theta) \right| \nonumber \\
&&= \frac{1}{c} \ {\rm exp} \left[- \frac{(M_{\rm ap}- M_0)}{{\rm c}}
\right].
\label{pdf}
\end{eqnarray}
Combining both probabilities, we obtain the likelihood function
\begin{equation}
{\rm L} (M_{\rm ap}^1, ..., M_{\rm ap}^{N_>}| p_0, c)  = P(N_>| N_{\rm
f},p_0)  \ \prod_{{\rm j}=1}^{N_>} \ \hat p(M_{\rm ap}^{\rm j}),
\label{likeli} 
\end{equation} 
which describes the probability to find $N_>$ aperture masses $>M_0$
where those above the threshold follow the distribution (\ref{pdf}).
From (\ref{likeli}) we can derive maximum likelihood estimates for
both fit parameters. Denoting these estimates by $(\hat c, \hat p_0)$
we obtain by differentiating (\ref{likeli}) w.r.t. $c$ and $p_0$ 
\begin{equation}
\hat c = \frac{1}{N_>} \ \sum_{j=1}^{N_>} \ M_{\rm ap}^{\rm j} - M_0
\label{c}
\end{equation}
and 
\begin{equation}
\hat p_0 = \frac{N_>}{N_{\rm f}};
\label{p}
\end{equation}
these results are of course not unexpected.
In order to show that (\ref{c}) and
(\ref{p}) are unbiased estimators we have to evaluate their ensemble
average which can be performed by applying the operator
\begin{eqnarray}
{\rm P} (X) &=&  f \sum_{N_>=1}^{N_{\rm f}} \ { N_{\rm f} \choose
N_> } \  p_0^{N_>} \ (1-p_0)^{N_{\rm f}-N_>}  \nonumber \\ 
&\times&\prod_{{\rm j}=1}^{N_>} \ \int_{M_0}^{\infty} {\rm d} M_{\rm
ap}^{\rm j} \ \hat p(M_{\rm ap}^{\rm j}) \ \ ( X ).
\label{op}
\end{eqnarray}
Averaging (\ref{c}) and (\ref{p}) with (\ref{op}) yields
\begin{equation}
\langle\hat c\rangle = {\rm P} (\hat c) = c \ {\rm and} \
\langle\hat p_0\rangle = {\rm P} (\hat p_0) = p_0 f.
\label{mest}
\end{equation}
Because of the second of eqs.(\ref{mest}), $\hat p_0$ is an unbiased
estimator for $p_0$  
only if $f \approx 1$. In the following we will consider only values of
$p_0$ and $N_{\rm f}$ which guarantee that $f \approx 1$ is satisfied,
otherwise the statistics is not good enough to determine the
parameters accurately anyway.
The correlation between the two estimators can be calculated from
\begin{equation}
\langle\hat c \ \hat p_0\rangle = {\rm P}  (\hat c \ \hat p_0) = c \ p_0 \ f.
\label{corr}
\end{equation}
Since we have $f \approx 1$ these two estimators are not correlated.
In the following we set $f=1$.
 
Dispersions of the estimators are defined by
\begin{equation}
\sigma_{\hat p_0} = \sqrt{\langle \hat p_0^2 \rangle -
\langle \hat p_0 \rangle^2}
\end{equation}
and 
\begin{equation}
\sigma_{\hat c} = \sqrt{\langle \hat c^2\rangle - \langle\hat
c\rangle^2}.
\end{equation}
The calculation of the dispersions involves the application of the
operator (\ref{op}) to the squared estimators.
After some algebra we obtain 
\begin{equation}
\langle\hat c^2\rangle = c^2 \ \left(1 + \sum_{N_>=1}^{N_{\rm f}} \ 
P(N_{\rm f},p_0) \ N_>^{-1}\right)
\end{equation}
and
\begin{equation}
\langle \hat p_0^2 \rangle = \frac{p_0}{N_{\rm f}} (1 - p_0) + p_0^2,
\end{equation}
which together with (\ref{mest}) leads to 
\begin{equation}
\sigma_{\hat c} =c \ \sqrt{\sum_{N_>=1}^{N_{\rm f}} \
P(N_>| N_{\rm f},p_0) \ N_>^{-1}},  
\label{disp_c}       
\end{equation}
and
\begin{equation}
\sigma_{\hat p_0}= 
\sqrt{\frac{p_0}{N_{\rm f}} (1 - p_0)}.
\label{disp_p0}
\end{equation}
As expected, the rms values of the estimators decrease with increasing
image size. This behaviour is obvious for (\ref{disp_p0}) and can be
seen for (\ref{disp_c}) if we use a recurrence relation to compute the
binomial distribution for a given $N_>$. The offset for the recurrence
relation, $C(1) = N_{\rm f} p_0 (1-p_0)^{N_{\rm f}-1}$, rapidly
decreases for large $N_{\rm f}$ and therefore the sum in (\ref{disp_c}).

Note that the rms values of the estimators depend only on their means
and $N_{\rm f}$ which is determined by the image size for a given
filter scale. Given that the mean values $(c, p_0)$ depend on the
filter radius $\theta$  and the threshold $M_0$ we can vary the
parameter triple $(\theta,M_0,N_{\rm f})$ to obtain a
significant difference between the various cosmological parameters.
Furthermore, redshift information coming from possible redshift
measurements of the haloes and the sources can be used to maximize the
difference between various cosmologies.

In Fig. \ref{multi_2_survey} we plot the mean values of the fit
parameters (denoted by crosses) for the filter radii $\theta=1,2,3,5$
arcmin for four cosmological models. The threshold $M_0$ is defined by
a signal-to-noise ratio of 5. For each cosmology and filter scale we
use a 25 deg$^2$ (thin lines) and 100 deg$^2$ (thick lines)
survey to compute the dispersions of
the estimators (\ref{p}) and (\ref{c}) which define the 1-$\sigma$
error ellipses in Fig. \ref{multi_2_survey}. As expected, if we double
the side length of the image the dispersions of $c$ and $p_0$ become
smaller by about a factor of two. The COBE normalized
EdS model is well separated for all filter scales. Since the
probability $p_0$ in this model is about a factor of 5 larger than
that of the remaining cosmologies we do not plot this model. 
The possibility of distinguishing
cosmologies strongly depends on the filter scale. For an aperture with
1 arcmin filter radius we can clearly distinguish the EdS(0.6,0.25)
and the low density model without cosmological constant from the other
cosmologies if we use the survey with the larger area. 
If we enlarge the filter radius, the differences between the 
two low-density models and the EdS model with large shape
parameter become smaller whereas the EdS(0.6,0.25) model remains
distinguishable from all the other cosmologies. 

In Fig. \ref{multi_mix_new} we demonstrate the effects on the
dispersions if we change the halo redshift integration range and the
source redshift $z_{\rm s}$, where we assume all sources to be located
at the same redshift. For reference, we choose the upper left plot
from Fig. \ref{multi_2_survey} with $L=10$ deg. If we shrink the halo
redshift interval to $z_{\rm d}=[0.6,1]$ we obtain the upper right
plot. Because of the stronger evolution of the halo number density for
high redshifts in the other cosmologies, the EdS(0.6,0.5) model can
now be distinguished from the low-density models. The EdS(0.6,0.25)
model has a value $p_0$ which is so small as to lead to a
significant deviation of $f$ -- as defined in (\ref{f}) -- from unity,
so that the expressions (\ref{disp_c}) and (\ref{disp_p0}) are no longer
valid, and therefore we have not plotted this model.

In the bottom left plot, compared  to the 
upper right one, we only consider haloes having redshifts in the
interval $z_{\rm d}=[0, 0.4]$. In this redshift range all cosmologies
considered are distinguishable. The reason for the low-density models
now being separated is the increasing difference between the rich
cluster mass functions of both models if we choose smaller redshifts.

In the remaining plot, in comparison with the top left
one, all sources are assumed to have the same redshift $z_{\rm s}=3$
which is about twice the mean source redshift used in the other panels.
Because of the improved efficiency of the weak lensing signal, the
probability is increased by about a factor of 4. For this very deep
survey all cosmologies are clearly separated.
From Figs. \ref{multi_2_survey} and \ref{multi_mix_new} it becomes clear
that we need large deep surveys in order to be able to distinguish
clearly between the
cosmological models considered, using only the statistics applied here.

If we want to compute $P(>M_{\rm ap},\theta)$ as expected
from real observations, we have to consider possible sources of error
which may change the theoretical value of $M_{\rm ap}$, because the
cross sections are given via aperture mass measurements (see
Fig. \ref{map_zeta}). To account for the intrinsic ellipticity
distribution of galaxies, we can proceed in the same way as for the
observable $N(>M_{\rm ap}, \theta)$ in KS99.  Essentially this means
that we convolve $P(>M_{\rm ap},\theta)$ with a Gaussian defined by
the dispersion of the intrinsic galaxy ellipticity distribution of the
sources.  As explained in SvWJK, this dispersion is expected to
dominate the noise in a measurement of $M_{\rm ap}$.
The probability obtained from the convolution can be directly compared
with observations. We have checked that, as expected from the results
of KS99, the convolution only slightly enhances the theoretical values
and does not change the shape of $P(>M_{\rm ap},\theta)$ appreciably
in the range of $M_{\rm ap}$ considered here. Therefore the fit
formula we obtained remains valid, and the maximum likelihood method can
be applied to derive dispersions for the fit parameters.

In Table \ref{table} we show the number density of haloes with
aperture mass larger than $M_{\rm ap}=0.04$ and cross-section radii
exceeding the threshold $\zeta_{\rm t}=0.8$ arcmin for the filter
radius $\theta=2$ arcmin. According to the fact that the rich cluster
mass function shows most clearly the cosmology dependence of the non-linear
evolution of the density field and since cross-section radii above 0.8
arcmin correspond to the most massive non-linear objects (see
Fig. \ref{zeta_m}), 
$N(>M_{\rm ap}, > 0.8', \theta)$ better distinguishes cosmologies than
$N(>M_{\rm ap}, \theta)$ for the same $M_{\rm ap}$ and $\theta$
(see KS99). A drawback when using the observable (\ref{numbercro}) is
the larger
image size required for the detection of significant differences
between various cosmologies. From the numbers in Table \ref{table}
we infer a survey area of 25 deg$^2$ which is needed to distinguish
the cosmologies considered here significantly, i.e., with no
overlapping Poissonian error bars. 

\begin{table}
\caption{ The number of haloes per square degree
                       with aperture mass larger 
                       than $M_{\rm ap}=0.04$ and cross-section radius
                       greater than $\zeta_{\rm t}=0.8$ arcmin, as
                       defined in (\ref{numbercro}), computed for five
                       cosmological models.
                       The  filter radius is $\theta=2$ arcmin.
          The aperture mass and the dispersion determined by the
	  filter radius correspond to a
          signal-to-noise ratio of 5.             
	}
\label{table}
\begin{tabular}{lccccc}
\noalign{\smallskip}\hline\noalign{\smallskip}
$\Omega_0$ & $\Lambda$ & $\Gamma$ & $\sigma_8$ & $N(> 0.8',> 0.04,2')$ \\
\noalign{\smallskip}\hline\noalign{\smallskip}
1.0       & 0 &      0.25 & 0.6 & 0.47    \\ 
1.0       & 0 &      0.25 & 1.0 & 11.9    \\
1.0       & 0 &      0.5  & 0.6 & 1.16    \\
0.3       & 0 &      0.25 & 1.0 & 2.22    \\
0.3       & 0.7 &    0.25 & 1.0 & 1.67    \\
\noalign{\smallskip}\hline\noalign{\smallskip}
\end{tabular}
\end{table}

\section{Discussion and Conclusions}

In this paper 
we computed the highly non-Gaussian tail of the 
probability distribution of the aperture mass $M_{\rm ap}$ resulting
from lensing by the large-scale structure. 
The CPDF is obtained by summing up the cross sections
of dark matter haloes with assumed spherical mass density 
following a universal NFW density profile.  
We used Press-Schechter theory to compute the number
density of massive haloes which cause the extended tail in the PDF of
$M_{\rm ap}$. 

The number density of haloes with large cross-sectional
radii, or in other words, with large masses, is a
sensitive measure for constraining the cosmological parameter set (see
KS99, and references therein). We showed that the number density in a
mass-selected sample of haloes with cross sectional radii above 0.8
arcmin, which corresponds to halo masses $\sim 10^{14} M_{\odot}/h$,
is measureable in all cosmologies considered. Especially, from a
deep, high-quality imaging survey of 25 deg$^2$, some of the currently
most popular cosmologies can be
distinguished by the varying number of haloes with the selected cross
section threshold. The expected number of massive haloes varies
from $\sim 12$ for the EdS(0.6,0.25) to $\sim 300$ for the high
normalised EdS model if we use the above-mentioned survey.

The main result of this work is an analytical formula describing the
PDF of the aperture mass, which turns out to be closely approximated by
an exponential for values of $M_{\rm ap}$ well above its rms. This fact
allows a simple maximum-likelihood analysis for distinguishing various
cosmological models, using the non-Gaussian tail of $M_{\rm ap}$
only, as demonstrated in Sect.\ 5.

Modifications of the present investigation can well be imagined and
may turn out to yield fruitful results in future. One is the use
of (photometric) redshift estimates which allow a more precise
measurement of the shear, owing to the redshift-dependent geometrical
factors entering the projected surface mass density. If the haloes
found by our method turn out to be associated with galaxy
concentrations, their redshift can be estimated, and the halo abundance as
a function of $M_{\rm ap}$ and redshift can be obtained, allowing to
greatly refine the statistics considered here and in KS99.

Our approach to attempt measuring cosmological parameters is
complementary to investigations employing two-point statistical
measures, such as the rms shear in (circular) apertures or the shear
two-point correlation function (e.g., Blandford et al. 1991; Kaiser
1992, 1998; Villumsen 1996; Jain \& Seljak 1997; Bartelmann \&
Schneider 1999), and those which consider the skewness of cosmic shear
statistics as a powerful handle on $\Omega_0$ (e.g., Bernardeau et al. 1997;
van Waerbeke et al. 1999; Jain et al. 1999). In contrast to these
lower-order statistical measures, the highly non-Gaussian features of
cosmic shear are expected to be less affected by noise, whose main
contribution is the intrinsic ellipticity dispersion of source
galaxies. 

We note that the cosmology dependence of our results are mainly (but
not exclusively) through the abundance of dark matter haloes as a
function of mass and redshift. Thus, our statistics provides a fairly
direct measure of this cluster abundance. In a future work, we shall
attempt to relate $M_{\rm ap}$ to the true three-dimensional mass of
haloes as determined in numerical $N$-body simulations, and thus
relate the aperture mass statistics directly to the mass function of
haloes. 

It should be stressed that, whereas the aperture mass is most likely
not the optimal statistics to measure the cosmic shear power spectrum
(see Kaiser 1998, and Seljak 1998, for different approaches), it is a
particularly convenient measure for highly non-Gaussian, spatially
localized features which can be obtained locally and directly from the
observed image ellipticities and whose noise properties can be
straightforwardly investigated. In particular, cosmic shear
measurement using $M_{\rm ap}$, and the search for mass-selected
haloes (Schneider 1996; KS99) are just two aspects of the same
underlying physics and statistics.

The validity of the various approximations which enter KS99 and the
current study needs to be investigated in more detail. In a
forthcoming paper (Reblinsky et al., in preparation) we will apply the
aperture mass statistics to the same numerical simulations used in
Jain et al.\ (1999), which combine very large N-body simulations with
ray-tracing methods. Such numerical simulations are indispensible not
only to assess the accuracy of the analytical approximation, but also to
study statistical estimators which cannot be calculated
analytically. Given the highly non-Gaussian nature of the projected
density field resulting from the evolved LSS density distribution, it
is by no means clear how to optimally and robustly distinguish between
different cosmological models. The use of the far tail of the $M_{\rm
ap}$-statistics should be viewed as one of several useful tools.

\section*{Acknowledgments}
This work was supported by the ``Sonderforschungsbereich 375-95 f\"ur
Astro-Teilchenphysik" der Deutschen Forschungsgemeinschaft. We want
to thank B. Geiger and L. van Waerbeke for many interesting
discussions, and M. Bartelmann for a careful reading of the manuscript.

\bsp

\label{lastpage}


\begin{thebibliography}{99}
\bibitem{bar} Bardeen, J.M., Bond, J.R., Kaiser, N. \& Szalay, A.S.
             1986, ApJ 304, 15
\bibitem{ba} Bartelmann, M. 1996, A\&A 313, 697
\bibitem{bs} Bartelmann, M. \& Schneider, P. 1999, A \& A, in press,
             astro-ph/9902152
\bibitem{bv} Bernardeau, F., van Waerbeke, L. \& Mellier, Y. 1997,
	     A\&A 322, 1
\bibitem{bl} Blandford, R.D. \& Jaroszy$\acute {\rm n}$ski, M. 1981, 
             ApJ 477, 27
\bibitem{bl1} Blandford, R.D., Saust, A.B., Brainerd, T.G. \& 
              Villumsen, J.V. 1991, MNRAS 251, 600
\bibitem{br} Brainerd, T.G., Blandford, R.D. \& Smail, I. 1996,
             ApJ 466, 623
\bibitem{ef} Efstathiou, G. 1996, in: {\it Cosmology and large scale
structure}, Les Houches Session LX, R. Scheffer, J. Silk, M. Spiro \&
J. Zinn-Justin (eds.), North-Holland, p.133.
\bibitem{gu} Gunn, J.E. 1967, ApJ 147, 61
\bibitem{ja} Jain, B. \& Seljak, U. 1997, ApJ 484, 560
\bibitem{ja1} Jain, B., Seljak, U. \& White, S.D.M. 1998, astro-ph/
              9804238
\bibitem{ja2} Jain, B., Seljak, U. \& White, S.D.M. 1999, astro-ph/
              9901191
\bibitem{ka1} Kaiser, N. 1992, ApJ 388, 272
\bibitem{ka} Kaiser, N. 1998, ApJ 498, 26
\bibitem{kr} Kruse, G. \& Schneider, P. 1999, MNRAS 302, 821 (KS99)
\bibitem{la} Lacey, C. \& Cole, S. 1993, MNRAS 262, 627
\bibitem{la2} Lacey, C. \& Cole, S. 1994, MNRAS 271, 676
\bibitem{me} Mellier, Y. 1998, preprint, To appear in Vol. 37 of 
             {\it Annual Reviews of Astronomy and Astrophysics}
\bibitem{me1} Mellier, Y., van Waerbeke, L. \& Bernardeau, F. 1998, preprint
\bibitem{na} Navarro, J.F, Frenk, C.S. \& White, S.D.M. 1996, 
             ApJ 462, 563 (NFW)
\bibitem{nav} Navarro, J.F, Frenk, C.S. \& White, S.D.M. 1997, 
             ApJ 490, 493 (NFW)
\bibitem{pr} Press, W.H. \& Schechter, P. 1974, ApJ 187, 425
\bibitem{re} Reblinsky, K., Kruse, G., Jain, B. \& Schneider, P. 1999,
             in preparation
\bibitem{sc} Schneider, P. 1996, MNRAS 283, 837
\bibitem{sc1} Schneider, P., van Waerbeke, L., Jain, B. \& Kruse, G.
              1998, MNRAS 296, 873 (SvWJK)
\bibitem{se} Seljak, U. 1998, ApJ 506, 64
\bibitem{vb} van Waerbeke, L., Bernardeau, F. \& Mellier, Y. 1999,
             A\&A 342, 15
\bibitem{vi} Villumsen, J. 1996, MNRAS 281, 369
\end{thebibliography}
\end{document}